\documentclass[preprint
 amsmath,amssymb,
 reprint,%
]{revtex4-2}
\usepackage{amsmath}
\usepackage{comment}
\usepackage{graphicx}
\usepackage{dcolumn}
\usepackage{bm}
\usepackage[utf8]{inputenc}
\usepackage[T1]{fontenc}
\usepackage{mathptmx}
\usepackage{etoolbox}
\usepackage[colorlinks=true]{hyperref}
\usepackage{nameref}
\usepackage[english]{babel}

\DeclareUnicodeCharacter{03C0}{\ensuremath{\pi}}      
\DeclareUnicodeCharacter{202F}{\,}                   
\DeclareUnicodeCharacter{2026}{\ldots}               

\makeatother
\begin{document}

\preprint{AIP/123-QED}

\title[Streamline controlled rectification of supercurrent in thin-film asymmetric weak links]{Streamline controlled rectification of supercurrent in thin-film asymmetric weak links}
\author{F. Antola}
\email{filippo.antola@sns.it}
\affiliation{ 
NEST Istituto Nanoscienze-CNR and Scuola Normale Superiore, I-56127 Pisa, Italy
}

\author{S. Battisti}
\affiliation{ 
NEST Istituto Nanoscienze-CNR and Scuola Normale Superiore, I-56127 Pisa, Italy
}

\author{A. Braggio}
\affiliation{ 
NEST Istituto Nanoscienze-CNR and Scuola Normale Superiore, I-56127 Pisa, Italy
}

\author{F. Giazotto}
\email{francesco.giazotto@sns.it}
\affiliation{ 
NEST Istituto Nanoscienze-CNR and Scuola Normale Superiore, I-56127 Pisa, Italy
}

\author{G. \surname{De Simoni}}
\email{giorgio.desimoni@nano.cnr.it}
\affiliation{ 
NEST Istituto Nanoscienze-CNR and Scuola Normale Superiore, I-56127 Pisa, Italy
}
\begin{abstract}
In this study, we examined the supercurrent diode effect (SDE) in mesoscopic superconducting weak links formed by asymmetric Dayem bridges. These planar metallic constrictions, which naturally exhibit Josephson-like behavior, offer a fundamental platform for investigating nonreciprocal transport phenomena in a regime where the bridge width aligns with the superconducting coherence length ($W \sim \xi$). The foundational concept is inspired by the Tesla valve, a classical fluidic device that achieves flow rectification through interference and turbulence between fluid streams enabled by geometric asymmetry. Analogously, we demonstrate that spatial asymmetry within superconducting structures can result in rectification due to the polarity-dependent interaction between transport and screening currents. By implementing controlled geometric defects at the junction between the constriction and superconducting leads, we induce current crowding and disrupt spatial inversion symmetry, thus facilitating directional switching behavior. Experimental results indicate a linear-in-field rectification regime at low magnetic fields, driven by the interaction between transport and screening currents, which is succeeded by complex vortex dynamics within the superconducting banks at elevated fields. Time-dependent Ginzburg–Landau simulations replicate significant features of the experimental observations and substantiate the influence of both screening currents and rearrangements of Abrikosov vortices. A comparative study across various geometries highlights the crucial role of defect shape and spatial confinement in determining the rectification efficiency, revealing a minimum threshold in bridge width below which crowding-induced SDE is significantly reduced. Our findings advocate for mesoscopic Dayem bridges as a flexible platform for designing and controlling superconducting diode functionalities.
\end{abstract}
\maketitle

\begin{figure}[t]
    \centering
    \includegraphics[width=\columnwidth]{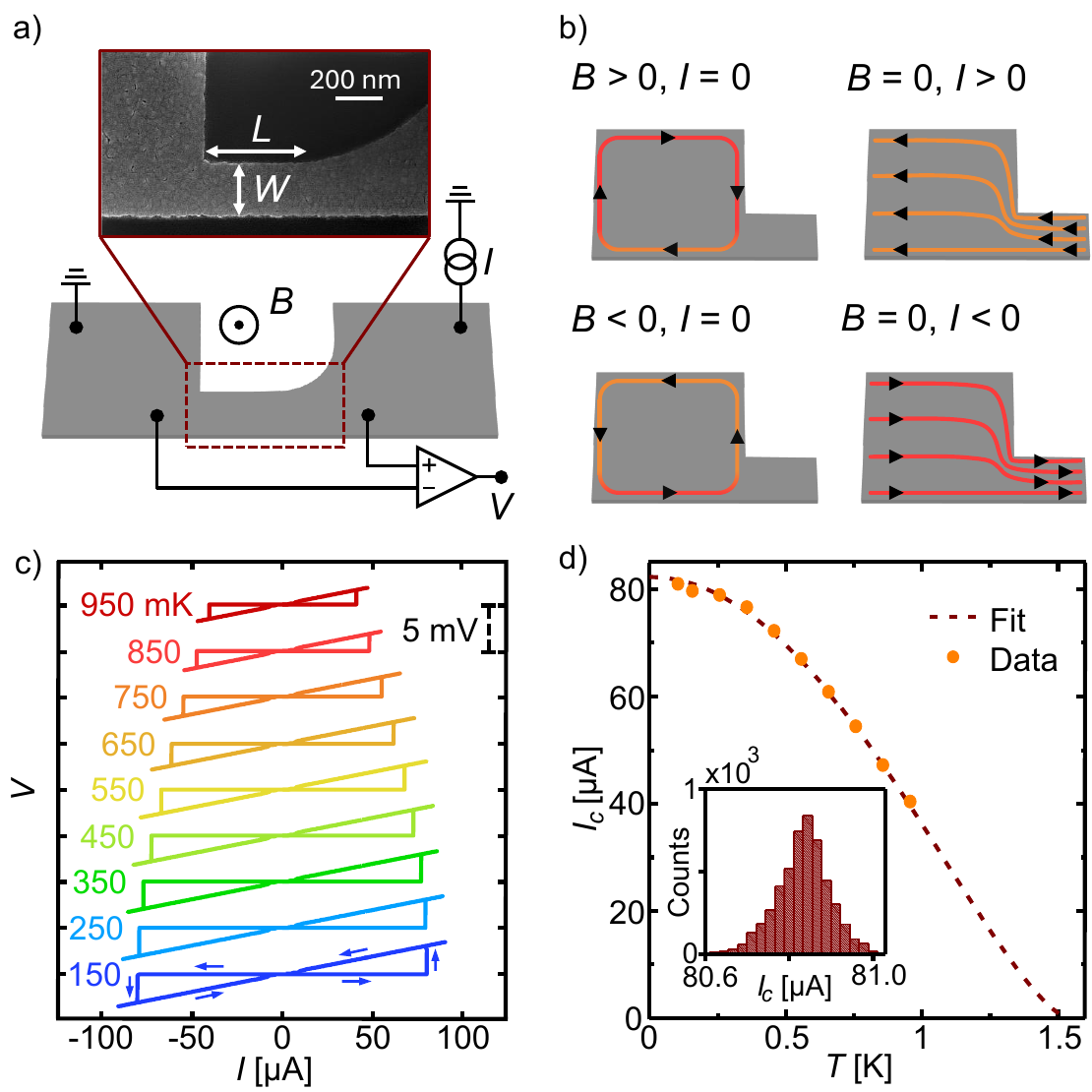}
    \caption{a) Schematic setup to measure the voltage characteristics $V$ as a function of a bias current $I$ with an applied out-of-plane magnetic field $B$. A close-up scanning electron micrograph (SEM) image of the bridge is also included. b) Left: schematic distribution of screening currents in the bridge and superconducting banks for opposite magnetic field polarities. Right: qualitative illustration of the current crowding effect at the geometric defect for opposite directions of the applied current. In both cases, the combined current density at the defect is enhanced or reduced depending on the relative orientation of transport and screening currents. The color indicates whether the screening currents locally add to the total current under positive (orange) or negative (red) bias. c) IV characteristics of the bridge for different bath temperatures (different colors). The temperature scale is expressed in mK. The curves are vertically offset by $5$ mV for clarity. The blue arrows indicate the cycle of the bias current swept back and forth, starting at zero amplitude. d) Temperature ($T$) dependence of the critical current $I_{C}$ for $B=0$. The red dashed line is the fit to the Bardeen equation. The inset shows the switching statistics at $T=100$ mK of a representative device.}
    \label{Fig 1}
\end{figure}
\section{Introduction}
A supercurrent diode (SD) is a superconducting device that supports a larger critical current in one direction than in the opposite direction, so that the positive ($I_C^+$) and negative ($I_C^-$) critical currents are different. 
Beyond scientific curiosity, the potential for groundbreaking technological applications serves as a significant driving force behind intense research on the supercurrent diode effect (SDE). Semiconductor diodes are indispensable components in modern electronics, enabling rectification, switching, and a multitude of other essential functions. The prospect of achieving a superconducting analog opens a pathway toward ultralow power electronics and enhanced energy efficiency in various applications based on nondissipative circuits, which could revolutionize computing~\cite{devoret_superconducting_2013, wendin_quantum_2017,braginski_superconductor_2019, hosur_digital_2024}, communication~\cite{mukhanov_superconductor_2004, aumentado_superconducting_2020}, signal processing~\cite{kleiner_superconducting_2004, irwin_transition-edge_2005} and power transmission~\cite{hassenzahl_electric_2004}.\\
The requirement for the SDE to manifest lies in the breaking of both spatial inversion and time-reversal symmetries within the superconducting system. The exploration of diverse strategies capable of inducing these symmetry breakings is central in the current research landscape in superconducting quantum technology. Several distinct physical mechanisms have been identified as potential origins of the SDE~\cite{nadeem_superconducting_2023, ma_superconducting_2025}. Inversion symmetry, which is naturally disrupted in non-centrosymmetric materials, can be broken in conventional superconductors with an ad-hoc nonreciprocal structure. In fact, the very first demonstrations of the superconducting diode effect utilized asymmetric heterostructured superconductors~\cite{ando_observation_2020, narita_field-free_2022}. Subsequently, various platforms have shown promising results, including proximitized two-dimensional electron systems~\cite{baumgartner_supercurrent_2022, costa_sign_2023,turini_josephson_2022,borgongino_biharmonic-drive_2025}, van der Waals heterostructures~\cite{wu_field-free_2022,bauriedl_supercurrent_2022,pal_josephson_2022,yun_magnetic_2023}, twisted multilayer graphene~\cite{lin_zero-field_2022,diez-merida_symmetry-broken_2023}. Finally, the application of an external magnetic flux is a convenient method to break the time-reversal symmetry to implement SDE in systems with specific geometries, such as chiral nanotubes~\cite{he_supercurrent_2023}, or interferometers~\cite{souto_josephson_2022, paolucci_gate-_2023,de_simoni_quasi-ideal_2024,greco_josephson_2023,greco_double_2024,chirolli_diode_2024,yerin_supercurrent_2024,yerin_supercurrent_2025}.
Among the techniques to induce the SDE by applying an external magnetic field, a method was identified even before the concept of the supercurrent diode was conceived. It exploits the combined action of an external magnetic field and a suitable asymmetry, constriction, or bend of the device. This idea was explored in thin films with wide channels~\cite{adami_current_2013,ilin_critical_2014}, where the transverse dimension $W$ is much larger than the coherence length $\xi$, and up to a certain extent, it represents the superconducting analogue of the fluidics Tesla valve. The latter is based on an array of baffles that converts part of the incoming viscous flow into a reverse backstream, obstructing propagation in one direction while allowing it in the other. Although such a mechanism cannot be reproduced in the Cooper pair condensate frictionless system, in these devices, the non-reciprocity arises because the stream of supercurrent flowing through the device and the screening current either add or subtract, depending on the position and the direction of the magnetic field. This results in an increased or suppressed supercurrent flow. Moreover, due to its geometry, the supercurrent density is enhanced in specific locations of the samples~\cite{clem_geometry-dependent_2011,clem_predicted_2012,hortensius_critical-current_2012,jonsson_current_2022} compared to what it would be in a perfectly straight wire. This mechanism is known as current crowding. Where the supercurrent is concentrated, the Bean–Livingston barrier~\cite{bean_surface_1964}, which regulates the nucleation of an Abrikosov vortex, is locally lowered~\cite{vodolazov_superconducting_2005}, making it easier for the vortex to form. Once inside, the vortex crosses the strip and annihilates with an antivortex at the opposite edge, producing a voltage drop that marks the onset of dissipation and defines the switching current. 
These locations, therefore, govern the device's switching behavior. Thus, the screening currents serve as a knob that allows the vortex nucleation energy to be lowered or raised at will at specific locations of the sample. Furthermore, since the sign of the bias and the screening current can be concordant or discordant, the switching current takes different values for positive and negative bias. This asymmetry reverses with the direction of the field. Finally, when the field is larger than lower critical field $B_{C1}$, the spatial configuration of the Abrikosov vortex pattern becomes also relevant, making the screening current paths further complex, and resulting in abrupt changes and peaks in the critical current vs. magnetic field characteristics~\cite{foltyn_probing_2023}. It is also worth noting that even minor fabrication imperfections or local inhomogeneities can induce current crowding and act as seeds for vortex nucleation. Indeed, SDEs have been observed even in nominally symmetric devices~\cite{suri_non-reciprocity_2022,hou_ubiquitous_2023,margineda_sign_2023}. Such effects are particularly relevant in the regime $W \gg \xi$, where current crowding scales with the ratio $W/\xi$ and small asymmetries are strongly amplified. Despite this complexity, several works have shown that in this regime, vortex pinning can be deliberately engineered to enhance SDEs~\cite{villegas_superconducting_2003,golod_demonstration_2022,lyu_superconducting_2021}.

Building upon these insights, here we introduce a family of SDs designed to investigate the interplay between current crowding and vortex dynamics in the mesoscopic regime in which the superconductor width is comparable to the superconducting coherence length, i.e., $W \sim \xi$. Devices incorporating such constrictions utilize Dayem bridge (DB) Josephson weak links, which are fundamental building blocks of superconducting electronics, offering a versatile platform. They feature a planar geometry compatible with standard nanofabrication and the ability to incorporate intentional geometric asymmetries.
Indeed, an intentional \textit{supercurrent rectification seed} is introduced at the edge of our DBs at the interface between one of the wide leads and the constriction, with the dual aim of breaking the spatial inversion symmetry and of inducing supercurrent crowding in a designated spot of the device. Although current crowding is a well-established rectification mechanism in macroscopic devices, its role remains largely unexplored in the confined regime, where the device is not large enough to host screening currents, thus behaving as a quasi-one-dimensional supercurrent channel. However, the micrometer-sized banks adjacent to the constriction provide a sufficient area to sustain magnetic screening currents, and at finite magnetic fields, they can host Abrikosov vortices. This introduces an additional level of complexity, enabling the investigation of the impact of vortex patterns and providing further insight into the mechanisms underlying nonreciprocal transport. In the following, we present a comparative analysis of different device configurations, which vary both the bridge width and the type of rectification seed. This strategy allows us to disentangle the respective contributions of confinement and asymmetry to the rectification efficiency.
\section{Engineering Nonreciprocity: Design and Characterization of the Asymmetric Bridge}
To properly design the device geometry, we first determined the normal-state resistivity of a reference aluminum film with thickness $t = 14\ \text{nm}$, obtaining $\rho \simeq 0.089\ \Omega \cdot \mu\text{m}$.
From this value, we estimate the superconducting coherence length using the standard expression for diffusive Bardeen-Cooper-Schrieffer superconductors:
$\xi = \sqrt{\frac{\hbar}{\rho N_F e^2 \Delta_0}} \simeq 80\ \text{nm}$,
where $N_F \simeq 2.15 \times 10^{47}\ \text{J}^{-1}\text{m}^{-3}$ is the density of states at the Fermi level for aluminum, and $\Delta_0 \simeq 230\ \mu\text{eV}$ the zero-temperature superconducting gap.
We also calculated the London penetration depth as $\lambda = \sqrt{\frac{\hbar \rho}{\pi \mu_0 \Delta_0}} \simeq 250\ \text{nm}$,
where $\mu_0$ is the vacuum permeability. The corresponding magnetic screening length is given by the Pearl length~\cite{pearl_current_1964,tinkham_introduction_2004}, $\Lambda = \frac{2\lambda^2}{t} \simeq 8\ \mu\text{m}$.
Since the film thickness $t$ is much smaller than both $\xi$ and $\lambda$, the system can be effectively treated as two-dimensional for the superconducting properties. For such a thin film, Al exhibits effective type-II behavior, despite being a type-I superconductor in the bulk. As a consequence, vortices can penetrate the film above the lower critical field $B_{C1}$.
Our devices are metallic weak links of uniform thickness connecting two superconducting banks~\cite{likharev_superconducting_1979} and feature a controlled geometric asymmetry. The aim is to induce non-reciprocal switching behavior by creating a localized region of current accumulation within the weak link. This is achieved by introducing a sharp-angled connection on one side of the bridge, which promotes current crowding, while the opposite side is shaped with a smooth transition. This geometric asymmetry enhances the local current density near the defect, allowing us to probe its influence on vortex formation and supercurrent rectification. 
The devices were fabricated by electron beam lithography (EBL) and a lift-off of $14$nm of Al deposited via e-beam evaporation. A detailed description of the fabrication procedure is reported in the \hyperref[sec:Appendix A]{Appendix A}.

We begin our discussion by focusing on a configuration that features a right-angled connection. A SEM image of a representative device, along with the measurement setup, is shown in Fig.~\ref{Fig 1}(a).
The width of the bridge was set to $W \simeq 200\ \text{nm} \sim 3\xi$, below the critical threshold $W_C \simeq 4.4\xi$ required to stabilize Abrikosov vortices within the constriction~\cite{likharev_superconducting_1979}. Although this size is sufficient to support spatial modulation of the current density (particularly at the entrances of the weak link, where the current crowding effect may take place), no screening currents are expected within the bridge itself. In other words, the supercurrent streamlines are all aligned in the same direction along the weak link, with possible variations in density but no reversal of flow. The spatial distribution of the screening currents is illustrated in the left panels of Fig.~\ref{Fig 1}(b). Conversely, the effect of current crowding at the geometric discontinuity is schematically depicted in the right panel of the same figure.
The bridge length was chosen as $L \simeq 400\ \text{nm} \approx 5\xi$, placing the device in the long-junction regime. This facilitates fabrication because slightly longer constrictions are more tolerant of lithographic imperfections. At the same time, the rectification mechanism is not expected to depend critically on $L$, and it should persist even at shorter junctions.
In both top and bottom sketches, the directions of transport and screening currents are aligned at the angle formed by the connection between the weak link and the lead, which acts as a seed for vortex nucleation and due to the supercurrent local density enhancement.
Figure~\ref{Fig 1}(c) shows the current–voltage (I–V) characteristics of a representative DB measured at selected bath temperatures in a DC-filtered dilution refrigerator. The device shows an abrupt transition to the normal state at the critical current $I_C$, and the characteristic resistance hystereticity typical of metallic weak links that arises from Joule heating in the bridge when the current is swept back from the resistive to the superconducting state~\cite{skocpol_selfheating_1974}.
The temperature dependence of $I_C$, shown in Fig.~\ref{Fig 1}(d), follows the Bardeen expression:
\begin{equation}\label{Bardeen}
I_C(T) = I_C(0) \left[1 - \left(\frac{T}{T_C}\right)^2\right]^{3/2},
\end{equation}
from which we extract a zero-temperature critical current $I_C(0) \simeq 82.3\ \mu\text{A}$ and a critical temperature $T_C \simeq 1.55\ \text{K}$. The latter allows us to estimate the zero-temperature superconducting energy gap as $\Delta_0 = 1.764\ k_B T_C \simeq 235\ \mu\text{eV}$, where $k_B$ is the Boltzmann constant. The good agreement with the Bardeen formula supports the picture of a weak link that is not deeply in the Josephson regime, but rather close to being a short wire.
The inset of Figure\ref{Fig 1}(d) shows an example of switching current statistics, which follows the characteristic asymmetric distribution typically observed in thermally activated switching events in superconducting constrictions~\cite{puglia_electrostatic_2020}.

\section{Rectification Behavior Under Magnetic Field}
\begin{figure}
    \centering
    \includegraphics[width=\columnwidth]{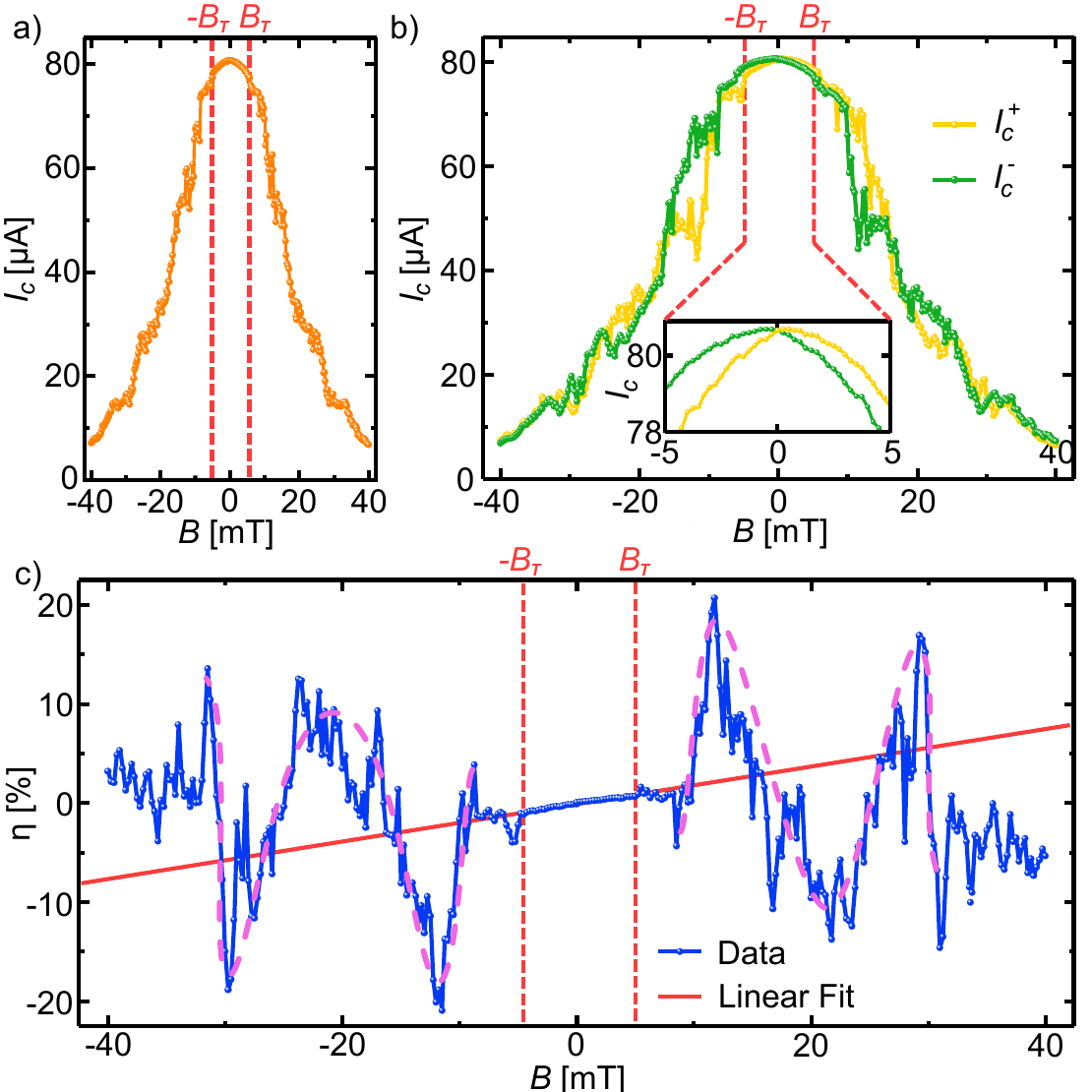}
    \caption{(a) Average absolute values of the critical currents in both directions as a function of the magnetic field $B$.  The red dashed lines indicate the linear rectification regime ($|B|<B_T$). (b) Behavior of the absolute values of the positive and negative critical current, $I_C^+$ and $I_C^-$, as a function of $B$. The inset displays a magnified view of the region corresponding to $|B|<B_T$. (c) Evolution of the rectification parameter $\eta$ vs. $B$. The dashed lines have the same meaning as in panel (a), while the dashed purple line serves as a guide to the eye. The red solid line represents the linear fit to the data in this regime.}
    \label{Fig 2}
\end{figure}

We now characterize the critical current $I_C=\frac{|I_C^+|+|I_C^-|}{2}$ upon the application of an out-of-plane magnetic field. Figure~\ref{Fig 2}(a) shows $I_C$ averaged over the switching current extracted by 100 acquisitions of the current-voltage characteristics measured by sweeping the bias current from 0 A to both positive and negative values. The associated error is given by the standard deviation and is smaller than the symbol size. The critical current decreases with increasing magnetic field and vanishes around $B \sim 40$ mT. The curve exhibits an even dependence on the magnetic field and reveals two distinct regimes. At low magnetic fields up to a threshold $B_T\sim5$ mT, the switching current decreases monotonically with $|B|$. For $|B|>B_T$, the trace becomes irregular and develops a series of abrupt jumps, which can be interpreted as due to the progressive entry of Abrikosov vortices into the type-II superconducting Al film above the first critical field $B_{C1}$. Rearrangement of the vortex pattern can alter the local current distribution and influence the switching behavior~\cite{foltyn_probing_2023}, resulting in the appearance of dips in the $I_C(B)$ curve.

Figure~\ref{Fig 2}(b) shows the magnetic field dependence of $I_C^+$ (in yellow) and $I_C^-$ (green) plotted separately. For $|B|> B_T$  both traces exhibit the same jump-like structure previously observed in panel (a), yet their evolution becomes strongly dependent on the current direction. Several crossing points emerge, corresponding to distinct intervals where $I_C^+$ or $I_C^-$ dominates. The direction dependence of the critical current, $I_C^+ \neq I_C^-$, is already visible for $|B|<B_T$, and becomes more pronounced as the field increases. This robust nonreciprocity over the entire field range confirms the occurrence of the SDE in our asymmetric weak link. The inset shows a magnified view of the low-field region ($|B| < B_T$), where it is possible to appreciate that the maxima of the two traces are shifted away from the zero field, occurring at $B^{\text{max}} \sim \pm 0.5\ \text{mT}$, with the sign determined by the current polarity. As a consequence, $I_C^+$ exceeds its zero-field value for $0 < B < B^{\text{max}}$, while $I_C^-$ does the same within $0 > B > -B^{\text{max}}$.

To quantify transport non-reciprocity, we introduce the conventional dimensionless diode efficiency parameter $\eta$, defined as: 
\begin{equation}\label{eta} 
\eta = \frac{I_{C}^+ - I_{C}^-}{I_{C}^+ + I_{C}^-},
\end{equation}
which yields a quantitative measure of the supercurrent rectification efficiency.
\noindent 
The evolution of the rectification parameter $\eta$ as a function of the magnetic field is shown in Fig.~\ref{Fig 2}(c). For $|B|<B_T$, $\eta$ increases linearly with the applied magnetic field, consistent with the monotonic behavior observed in $I_C^\pm$. A linear fit in this region yields a slope of $m_R^{\text{EXP}} \sim 0.17\%\ \text{mT}^{-1}$.
As the magnetic field increases over $B_T$, $\eta$ exhibits a distinctly non-monotonic trend, with rectification peaks and valleys reaching values as high as $20\%$. Due to the multiple crossings of $I_c^+$ and $I_C^-$, $\eta$ reverses its sign several times, realizing a low-frequency, quasi-periodic oscillation, as highlighted by the pink dashed guide for the eye in Fig.~\ref{Fig 2}(c). This pattern is superimposed on sharper variations and abrupt jumps, which reflect the irregular switching behavior at higher fields. In particular, the $\eta(B)$ curve displays an approximately overall odd symmetry to the magnetic field, i.e., $\eta(-B) = -\eta(B)$. 
We believe that this is a distinctive signature of the physical mechanism underlying supercurrent rectification, which can be inferred from the current distributions sketched in Fig.~\ref{Fig 1}(b). 
Although the bridge itself is too narrow to support the screening, the screening currents hosted by the lead may still influence the local current distribution near the seed angle, where the transport current is crowded. Depending on the direction of the applied current, the screening currents could enhance or suppress the local current density at that point. Since the intensity of screening currents increases approximately linearly for a weak applied magnetic field~\cite{clem_predicted_2012}, this scenario may account for the initial linear growth of $\eta$ observed at fields below $B_T$.
At higher magnetic fields, \textit {i. e.}  for $|B|>B_T$, the system enters a more intricate behavior in which current crowding interplays with the Abrikosov vortex screening pattern. This regime cannot be intuitively captured by a simple qualitative sketch of the screening currents, requiring instead a more accurate model that also incorporates the dynamics of nucleation and pinning of Abrikosov vortices in the 2D regions adjacent to the bridge. These rectification sources are intrinsically odd with respect to the magnetic field, as reflected in the general antisymmetric behavior of \(\eta(B)\). As a consequence, separating the odd and even components of the rectification signal may serve as an effective tool to discriminate genuine rectification effects from noise contributions. 
A discussion of this approach is provided in \hyperref[sec:Appendix B]{Appendix B}.

\section{Time-dependent Ginzburg–Landau model}
\begin{figure}
    \centering
    \includegraphics[width=\columnwidth]{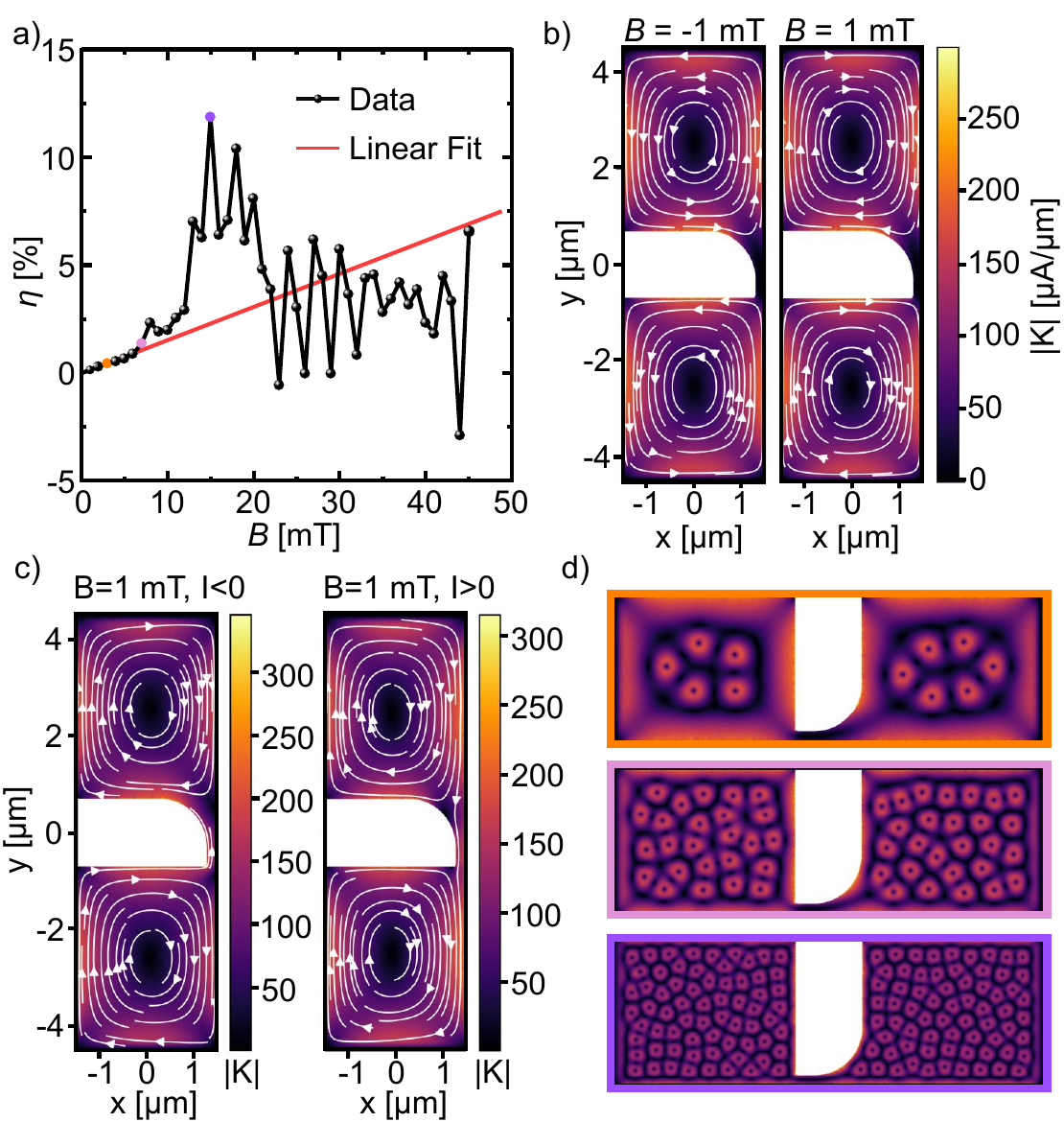}
    \caption{(a) $\eta$ vs $B$ calculated using the TDGL formalism. (b) Spatial distribution of the screening current density $\mathbf{K}$ in the superconducting banks for opposite values of the external magnetic field. (c) Streamlines of the total sheet current density $\mathbf{K}$ in the device under opposite bias directions at a fixed magnetic field. The scale for $\mathbf{K}$ is the same as (b). (d) Vortex configurations at selected magnetic field values. The color of each frame matches the color of a corresponding point in panel (a).}
    \label{Fig 3}
\end{figure}

To gain insight into the rectification mechanism, we performed numerical simulations based on the time-dependent Ginzburg–Landau (TDGL) model, which allowed us to calculate the time and spatial resolved values of the superconducting order parameter $\psi$, of the sheet current density $\mathbf{K}$, and of the electric potential $\mu$, as function of the device bias current and the external magnetic field $B$. In particular, we numerically solved the TDGL equations on a two-dimensional unstructured mesh, in dimensionless form, using experimentally determined values of $\xi$ and $\lambda$ to scale the physical quantities.  We computed the voltage response of the system as a function of the applied current bias. A threshold on the time-averaged voltage drop across the device was then used to extract the critical current for both bias directions over a range of magnetic fields. The resulting rectification parameter $\eta$ is shown in Fig.~\ref{Fig 3}(a). A detailed description of the model is reported in \hyperref[sec:Appendix C]{Appendix C}.
The simulation shows good quantitative agreement with the experimental data in the initial linear regime, where a linear fit yields a slope of $m_R^{\text{SIM}} \simeq 0.18\%\  \text{mT}^{-1}$, consistent with the experimental value. Given the ideal nature of the model, it does not account for the intrinsic granularity of the system, which becomes particularly relevant at high magnetic fields, where defects can act as pinning centers. For this reason, we did not further refine the magnetic field sampling, as the model itself is not suited to capture disorder-driven effects accurately. Still, the TDGL calculations reproduce the essential physics, including the abrupt oscillations and sign changes of $\eta$.
Figure~\ref{Fig 3}(b) shows the simulated current density distribution $\mathbf{K}$ under an applied magnetic field, in the absence of transport current. This panel highlights the system's screening response and confirms that the screening currents circulate only within the superconducting banks, with no significant flow through the narrow bridge. This supports our earlier assumption that the bridge itself cannot support screening due to its reduced width and effective dimensionality. Figure~\ref{Fig 3}(c) displays the streamlines of $\mathbf{K}$ in the presence of both screening and transport currents, for opposite bias directions. When a negative current is applied (from bottom to top), the bias and screening currents combine near the right-angle corner of the bridge, locally enhancing the current density. In contrast, for a positive current, the screening currents flow in the opposite direction, partially compensating the bias current and reducing the local density. This polarity-dependent enhancement or suppression of current density in the crowded region demonstrates the directional nature of the crowding effect and its central role in allowing rectification for $|B|<B_T$.
Figure~\ref{Fig 3}(d) shows the spatial distribution of vortices at various magnetic field values. As can be seen in the top-left panel, vortices are already present under $B_T$, indicating that magnetic flux penetration begins at relatively low fields. The estimated upper bound for the first critical field $B_{C1}<3 \ \text{mT}$, corresponding to the onset of vortex entry, lies within the expected range between the bulk value~\cite{tinkham_introduction_2004} ($\sim 3\ \text{mT}$) and the two-dimensional limit~\cite{stan_critical_2004,kubo_tuning_2023} ($\sim 1\ \text{mT}$), as determined from our material parameters. Here, a sparse Abrikosov lattice forms, with vortices located primarily at the center of the superconducting banks and well separated from the constriction. As the magnetic field increases, the number of vortices grows and the lattice expands, gradually approaching the bridge region. Eventually, some vortices reach the vicinity of the current-crowded area near the geometric defect, where their presence is likely to perturb the local current distribution and influence the rectification response. This progressive interaction between the evolving vortex configuration and the bridge geometry accounts for the breakdown of the linear behavior of $\eta$ and the emergence of non-monotonic rectification in higher fields. 
In summary, the TDGL simulations support the interpretation of the rectification mechanism in terms of current crowding and screening effects for $|B|<B_T$, which account for the linear increase observed in $\eta$ for weak magnetic fields. At higher values, the evolution of the vortex configuration and its interaction with the current-crowded region introduce additional complexity, leading to a breakdown of linearity and the emergence of sign reversals in the rectification signal.
\begin{figure*}
    \centering
    \includegraphics[width=\textwidth]{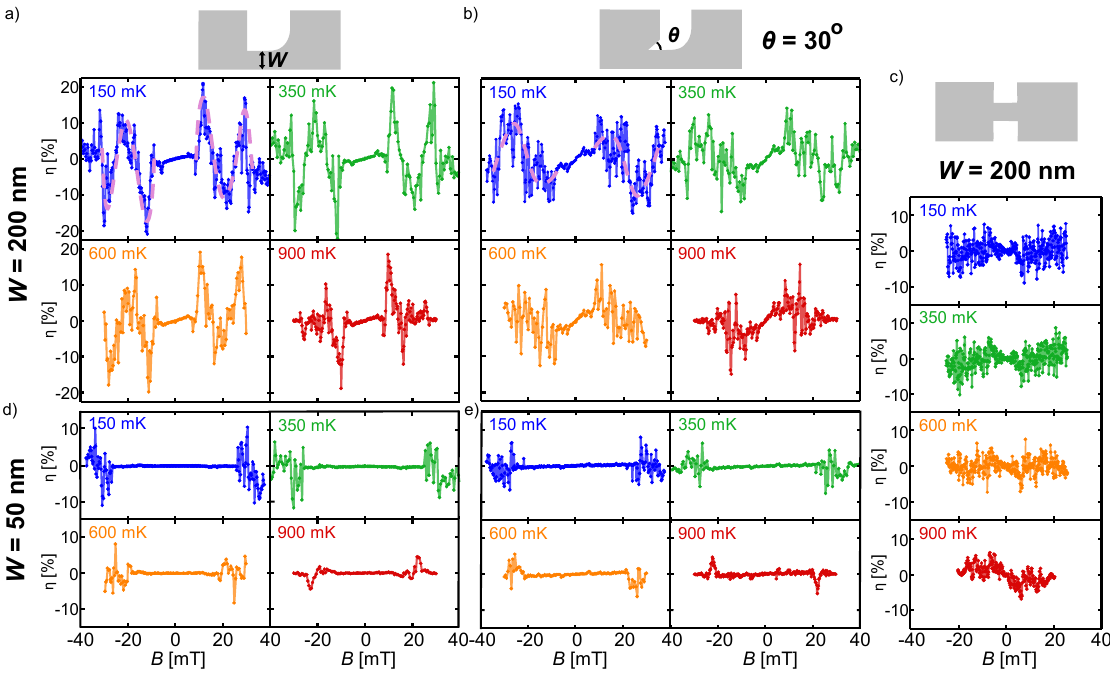}
    \caption{ $\eta$ vs $B$ for different thermal bath temperatures for devices with a bridge width of 200 nm, featuring a right-angled defect (a), a sharp-angled defect (b), and a symmetric geometry (c). Sketches above each panel illustrate the corresponding device geometry. The same applies to (d) and (e), but for a bridge with a width of 50 nm and a right-angle (d) and a sharp angle (e). The violet curve in (a) and (b) serves as a guide to the eye.}
    \label{Fig 4}
\end{figure*}
\section{Temperature and Geometry Dependence}
\noindent
In this section, we systematically examine how the temperature and shape of the rectification seed affect the response of the supercurrent diode. Figure~\ref{Fig 4} offers a comparative overview of several device configurations. To assess the role of spatial confinement, we compare two bridge widths: a wide geometry ($W = 200$ nm) and a narrow geometry ($W = 50$ nm). Furthermore, we investigate the effect of geometric asymmetry by analyzing two types of defects: a right-angled connection and a sharper constriction with a nominal angle of $30^\circ$. For reference, we also include a fully symmetric device that features identical right-angled connections on both sides of the bridge.

Panel (a) shows the rectification behavior of the same device studied previously, now measured at four different temperatures. Although in the linear regime, it is essentially unaffected throughout the temperature range, for $|B|>B_T$ it becomes progressively more complex: it remains qualitatively similar up to 600 mK, while at 900 mK only the initial rectification peak persists (with nearly unchanged amplitude), followed by a noisier and more irregular signal. It is worth highlighting that the described behavior at 900 mK more closely resembles the simulation results in Fig.~\ref{Fig 3}(a), consistent with the validity of the Ginzburg-Landau theory being formally limited to $T\sim T_C$. 
Panel (b) explores the effect of geometric asymmetry by replacing the right-angled defect with a sharper constriction forming a nominal $30^\circ$ angle. The rectification efficiency under $B_T$ increases significantly, with a linear fit producing $m_{\text{SH}}^{\text{EXP}} \simeq 0.73\%\ \text{ mT}^{-1}$. This enhancement is consistent with a major role for current crowding, which is expected to be stronger for sharper angles~\cite{clem_geometry-dependent_2011}. At higher magnetic fields, the rectification response exhibits the same qualitative features as observed in the right-angle case. However, the evolution of $\eta$ with flux is more chaotic, resulting in a strong instability of the rectification properties. As the temperature increases, the \textit{ slowly evolving} pattern is progressively suppressed, and only the low-field linear regime remains identifiable.

Panel (c) shows the response of a fully symmetric geometry, in which a bridge with the same width as in previous configurations forms two identical right-angled connections on both sides, symmetrically linking it to the electrodes. In this configuration, both the current-crowding regions and the surrounding banks are nominally symmetric, and no net rectification is expected.
However, the experimental data reveal a highly chaotic $\eta$ pattern, likely resulting from competition between nearly equivalent crowding sites or from competing vortex nucleation seeds originating from unwanted defects in random locations of the device. Indeed, in the presence of minor fabrication imperfections, this symmetry renders the system extremely sensitive to magnetic field fluctuations, resulting in erratic rectification behavior.

Panels (d) and (e) present devices with the same right-angled and sharp-angled defects discussed previously, but with a reduced bridge width of $W \simeq 50$ nm, comparable to the superconducting coherence length $\xi$. These narrow, quasi-one-dimensional channels exhibit negligible rectification up to magnetic fields as high as $B_T^{50 nm} \simeq 20$ mT. In this regime, suppression of the critical current due to current crowding, which scales approximately as $(\xi/W)^n$, with a geometry-dependent exponent $n$~\cite{clem_geometry-dependent_2011}, becomes ineffective, highlighting the key role of geometric confinement in enabling the diode effect. At higher magnetic fields, a high-frequency rectification pattern emerges in both devices, reaching amplitudes of up to 10\%. As the temperature increases, this complex structure smooths out into a well-defined single peak of approximately $5\%$, with opposite polarity in the two geometries.

\section{Conclusions}
We have shown that asymmetric Dayem bridges provide a minimal yet powerful platform for investigating the superconducting diode effect (SDE). Much like a classical Tesla valve, where fluid streams are guided to combine or cancel based on flow direction, these systems rely on constructive or destructive interference between transport and screening currents, depending on the bias polarity. At low magnetic fields, this interaction gives rise to a linear nonreciprocal response. As the field increases, Abrikosov vortex dynamics in the superconducting banks become dominant, producing quasiperiodic features and multiple sign reversals in the rectification signal.
Our experimental results, supported by time-dependent Ginzburg–Landau simulations, reveal rectification efficiencies reaching up to $20\%$. We identify key design parameters, such as defect sharpness and spatial confinement, that govern both the magnitude and the robustness of the effect. Notably, even nominally symmetric devices exhibit rectification, highlighting the high sensitivity of the SDE to minute, fabrication-induced asymmetries, which intrinsically limit the device performance tunability. In contrast, rectification vanishes as the bridge width approaches the superconducting coherence length, establishing a lower bound for diode functionality. In our devices, the large superconducting banks can host multiple vortices, which likely contribute to the complexity and irregularity of the high-field response. A promising strategy to improve rectification control could involve designing leads that allow for the deterministic placement of a limited number of vortices, either through tailored geometries or engineered pinning sites. 
This geometry-driven approach offers a practical strategy for realizing compact, low-dissipation, direction-sensitive components for next-generation superconducting quantum and cryogenic electronics.
\section*{Acknowledgments} \label{sec:acknowledgements}
We acknowledge the EU’s Horizon 2020 Research
and Innovation Framework Programme under Grant No. 101057977 (SPECTRUM)
and the PNRR MUR project PE0000023-NQSTI for partial
financial support. A.B. acknowledges the 
MUR-PRIN 2022—Grant No. 2022B9P8LN-(PE3)-Project NEThEQS “Non-equilibrium coherent thermal effects in quantum systems” in PNRR Mission 4-Component 2-Investment 1.1 “Fondo per il Programma Nazionale di Ricerca e Progetti di Rilevante Interesse Nazionale (PRIN)” funded by the European Union-Next Generation EU and CNR project QTHERMONANO.
\appendix
\section{Device Fabrication and Electrical Characterization}\label{sec:Appendix A}
The devices consist of a 14 nm-thick Al film fabricated by a single-step electron-beam lithography of poly methyl methacrylate (PMMA) resist mask and evaporation of metal onto an oxidized silicon wafer (the SiO$_2$ is 300 nm thick) in an ultra-high vacuum electron-beam evaporator with a base pressure of about 10$^{-11}$ Torr. The latter was carried out at a rate of 0.05 nm/s, followed by a standard lift-off process in acetone. 

The switching current was identified as the point at which the device transitioned from the superconducting to the resistive state. Electrical characterization of the asymmetric weak links was performed using a four-wire measurement technique in a filtered $^3$He-$^4$He  dilution refrigerator. A low-noise current bias was applied, and the voltage drop across the weak links was measured using a room-temperature DLPVA-101-F-D differential preamplifier and a National Instruments NIDAQx USB-6002 digital-to-analog and analog-to-digital board.
\section{Symmetry-based analysis of the rectification signal}\label{sec:Appendix B}
We present a symmetry-based decomposition of the rectification signal $\eta(B)$ into its odd and even components with respect to the applied magnetic field $B$. This method provides an effective tool for discriminating genuine rectification effects, which are expected to be odd functions of $B$, from possible noise artifacts or spurious contributions that are even.
The odd and even components of $\eta(B)$ are defined as:
\begin{equation}
\eta_{\mathrm{odd}}(B) = \frac{\eta(B) - \eta(-B)}{2}
\end{equation}
\begin{equation}
\eta_{\mathrm{even}}(B) = \frac{\eta(B) + \eta(-B)}{2}
\end{equation}
\begin{figure*}
    \centering
    \includegraphics[width=\textwidth]{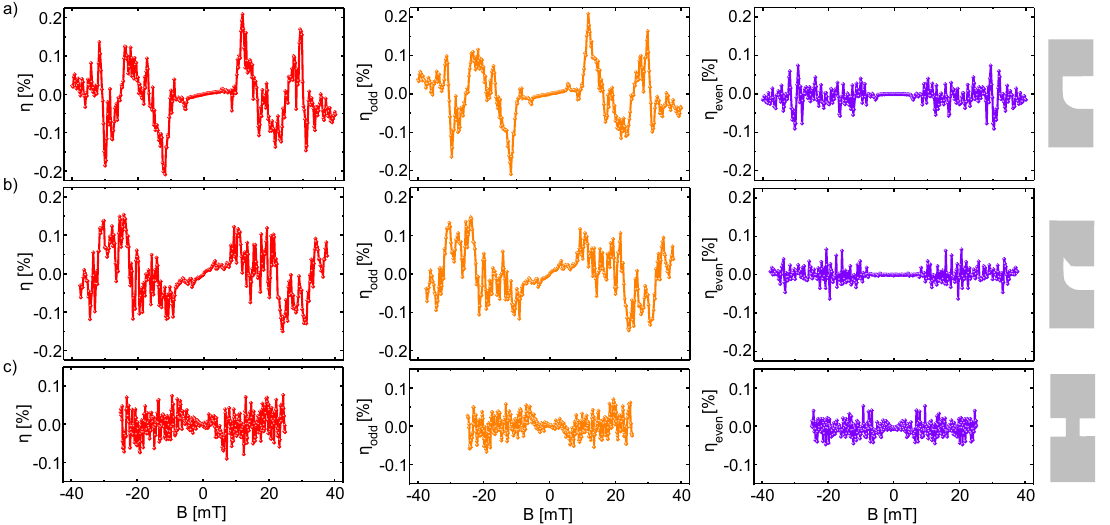}
    \caption{Symmetry decomposition of $\eta(B)$ (red) into its odd component $\eta_{\mathrm{odd}}$ (orange) and even $\eta_{\mathrm{even}}$ (violet) for the right angle (a), sharp angle (b) and symmetric configuration (c). All measurements are performed at 150 mK.}
    \label{Fig 5}
\end{figure*}
We apply this procedure to the three devices featuring a $200$ nm bridge width, namely the right-angle, the sharp-angle, and the symmetric configurations. The rectification signals, measured at $150$ mK, together with their odd and even components, are reported in Fig.~\ref{Fig 5}.\\
In Fig.~\ref{Fig 5}(a) and \ref{Fig 5}(b), the asymmetric devices, each with a single current crowding source, show an odd component $\eta_{\mathrm{odd}}(B)$ that closely matches the behavior of the measured rectification signal, confirming its origin. The even component $\eta_{\mathrm{even}}(B)$ is strongly suppressed in the linear regime ($|B| < B_T$), further validating the robustness of the signal against spurious contributions. At higher magnetic fields ($|B| > B_T$), it becomes more pronounced, likely due to an increased influence of flux noise, since the device becomes more sensitive to field fluctuations, or to additional rectification mechanisms beyond our model. In contrast, in the symmetric configuration of Fig.~\ref{Fig 5}(c), the odd and even components are comparable. This behavior reflects the system's enhanced sensitivity to flux fluctuations, as no specific site exists where the switching is preferentially triggered, unlike in the asymmetric devices.
\section{TDGL Model}\label{sec:Appendix C}
Numerical simulations were carried out using the time-dependent Ginzburg-Landau (TDGL) formalism as implemented in the open-source \textit{pyTDGL} package~\cite{bishop-van_horn_pytdgl_2023}, which solves the generalized TDGL equations for thin-film superconductors with arbitrary two-dimensional geometries.
The superconducting order parameter $\psi$ evolves according to the dimensionless TDGL equation:
\begin{align}\label{GL}
    &\frac{u}{\sqrt{1+\gamma^2|\psi|^2}}\left(\frac{\partial}{\partial t}+i\mu+\frac{\gamma^2}{2}\frac{\partial |\psi|^2}{\partial t}\right)\psi =\\&= (1-|\psi|^2)\psi+(\nabla - i\mathbf{A})^2\psi,
\end{align}
where $\mu(\mathbf{r},t)$ is the scalar electric potential and the magnetic field $B$ is introduced through the vector potential $\mathbf{A}(x,y)$, defined using the symmetric gauge:
\begin{equation}
\mathbf{A}(x,y) = -\frac{By}{2}\hat{\mathbf{x}} + \frac{Bx}{2}\hat{\mathbf{y}}.
\end{equation}
Given that $W \ll \Lambda$, self-induced fields can be neglected~\cite{heron_hierarchy_2003}, and only the external magnetic field is considered. The parameters used are $u \simeq 5.79$, which characterizes the ratio of amplitude to phase relaxation times in dirty superconductors, and $\gamma \simeq 5$~\cite{foltyn_probing_2023}, which accounts for inelastic scattering.
The total current density is given by the sum of the normal- and super-current $\mathbf{J}_N$ and $\mathbf{J}_S$:
\begin{equation}
\mathbf{J} = \text{Im}[\psi^*(\nabla - i\mathbf{A})\psi] - \nabla \mu = \mathbf{J}_S + \mathbf{J}_N,
\end{equation}
and, for thin-film systems with $t \ll \lambda$, we compute the sheet current density as $\mathbf{K} = t(\mathbf{J}_S + \mathbf{J}_N)$. The potential $\mu$ is obtained by solving the Poisson-like equation derived from current conservation:
\begin{equation}
\nabla^2\mu = \nabla \cdot \text{Im}[\psi^*(\nabla - i\mathbf{A})\psi].
\end{equation}
All quantities were expressed in dimensionless units, normalized on the experimentally determined values of $\xi$ and $\lambda$. The simulations were performed on a two-dimensional unstructured mesh, ensuring sufficient resolution to capture variations in both $\psi$ and the current density.

%

\end{document}